\documentclass{article}
\usepackage{amssymb}

\input{tcilatex}
\begin{document}

\subsection{\protect\bigskip A Theory for Spin Glass Phenomena in
Interacting Nanoparticle Systems}

\begin{center}
\bigskip \bigskip

Derek Walton\textbf{\medskip }

Dept. of Physics and Astronomy, McMaster University\medskip 

\bigskip 
\end{center}

\bigskip 

Dilute magnetic nanoparticle systems exhibit slow dynamics [1] due to a
broad distribution of relaxation times that can be traced to a
correspondingly broad distribution of particle sizes [1]. However, at higher
concentrations interparticle interactions lead to a  slow dynamics that is
qualitatively indistinguishable from that dislayed by atomic spin glasses.\
A theory is derived below that accounts quantitatively for the spin-glass
behaviour. The theory predicts that if the interactions become too strong
the spin glass behaviour disappears. This conclusion is in agreement with
preliminary experimental results. 

\bigskip 

At high temperatures assemblies of nanoparticles are superparamagnetic, but
on cooling it has been shown [2] that they enter a spin glass phase. \ Dc
magnetic relaxation\ has revealed ageing effects [3], and similar results
have been obtained for the relaxation of the low-frequency ac susceptibility
[4]. While ageing phenomena are certainly characteristic of spin glasses
they are also displayed by many random systems, what possibly sets spin
glasses apart is a dynamic memory effect that has been referred to as
rejuvenation [5,6,7]: if the cooling in zero field is halted the
susceptibility slowly decays. On resuming the cooling the susceptibility
recovers to the value it would have had if the cooling had not been
interrupted. On subsequent heating a decreased susceptiblity is observed in
the vicinity of the temperature at which the cooling was halted. These
effects have been observed in interacting nanoparticle systems [8,9,10,11]

\bigskip \bigskip 

The most elementary theory for magnetic interactions is a mean field theory;
so in this spirit, the model that will be studied considers that a local
interaction field is produced by the local magnetization that affects the
local susceptibility. Because of the disorder, the local magnetization is
spatially restricted, but the correlation length of the local field at low
temperatures can be expected to be much larger than the average particle
separation. At high temperatures all the particles are able to achieve their
equilibrium distribution, but as the material is cooled the relaxation times
increase, and eventually when the relaxation time of a particle becomes so
long that it is unable to relax in the time remaining in the cooling process
it becomes blocked. After cooling in zero field the magnetization of the
sample must be zero. If a small field is applied and the sample slowly
heated from a temperature that is low enough for all the particles to be
blocked, some of the blocked grains will be remagnetized, and the sample
will begin to acquire a moment. Initially the smallest particles with the
shortest relaxation times are remagnetized followed by the larger particles
as the temperature increases. The particle size distribution for these
materials displays a broad peak, and is often well approximated by a
log-normal distribution; so the number and size of remagnetized particles
increases rapidly with temperature. The net magnetization is the result not
only of the preferential orientation of the particle moment imposed by the
field, but also the preferential growth of the correlation length for the
local magnetization. The first is much larger than the second; so the
increase in correlation length will be neglected.

\bigskip

If the cooling is interrupted, and the sample held at an intermediate
temperature, $T_{w}$ for a time $t_{w}$, some of the blocked particles,
whose number increases \ as $\ln t_{w}$\ [1]\ can equilibrate at $T_{w}$.
This changes the local field. On resuming cooling particles whose blocking
temperature is $\leq T_{w}$ will again be blocked, but now a small number of
particles, whose blocking temperature was greater than $T_{w}$ will have the
equilibrium distribution at $T_{w}$, and the susceptibility will be slightly
changed. On heating in a small field there will be a small difference in the
magnetization. Initially this difference will be negligible because only the
smallest particles, that make the least contribution to the moment, will
become unblocked at low temperatures. However on further heating larger and
more numerous grains become unblocked, and the moment difference increases
rapidly until $T_{w}$ is reached. At this temperature the grains that were
responsible for the effect of waiting on the susceptibility begin to be
remagnetized, and the moment difference rapidly disappears.

With a mean field theory, the local field $H_{l}$ is proportional to the
local magnetization density, $M_{l},$ $H_{l}=\lambda M_{l}$. In the present
case assemblies of nanoparticles have a range of grain sizes, and $%
H_{l}=\lambda \int dVN(V)m_{l}(V)$, where $N(V)$ is the normalized grain
size distribution, and $m_{l}(V)$ is the orientationally averaged
magnetization of a grain whose volume is $V$ .

\bigskip

Consider a subset of particles of volume $V$ , all making an angle $\theta $
with $H_{l}$, their irregular shape leads to a single easy axis for their
magnetiztion; so there are two possible orientations of their magnetic
vectors, and the approach to equilibrium is governed by the master equations:

\begin{center}
$\frac{dn^{+}}{dt}=w^{-}n^{-}-w^{+}n^{+}=n^{-}\omega e^{-\frac{%
KV+JH_{l}V\cos \theta }{T}}-n^{+}\omega e^{-\frac{KV-JH_{l}V\cos \theta }{T}%
} $

$\frac{dn^{-}}{dt}=w^{+}n^{+}-w^{-}n^{-}=n^{+}\omega e^{-\frac{%
KV-JH_{l}V\cos \theta }{T}}-n^{-}\omega e^{-\frac{KV+JH_{l}V\cos \theta }{T}%
} $
\end{center}

where $\omega $ is an attempt frequency on the order of $10^{8}$ Hz, $K$ is
the anisotropy constant, $J$ is the saturation magnetization, and all
energies are in temperature units. Subtracting,

\begin{center}
\bigskip

$\frac{dn^{+}}{dt}-\frac{dn^{-}}{dt}=2\omega e^{-\frac{KV}{T}}(n^{-}e^{-%
\frac{JH_{l}V\cos \theta }{T}}-n^{+}e^{\frac{JH_{l}V\cos \theta }{T}})$

$=\omega e^{-\frac{KV}{T}}[(n^{-}e^{-\frac{JH_{l}V\cos \theta }{T}}-n^{+}e^{%
\frac{JH_{l}V\cos \theta }{T}})+(n^{-}e^{\frac{JH_{l}V\cos \theta }{T}%
}-n^{+}e^{-\frac{JH_{l}V\cos \theta }{T}})+(n^{-}e^{-\frac{JH_{l}V\cos
\theta }{T}}-n^{+}e^{\frac{JH_{l}V\cos \theta }{T}})-(n^{-}e^{\frac{%
JH_{l}V\cos \theta }{T}}-n^{+}e^{-\frac{JH_{l}V\cos \theta }{T}})]$

$=\omega e^{-\frac{KV}{T}}[(n^{-}-n^{+})2\cosh \frac{JH_{l}V\cos \theta }{T}%
-2n^{-}\sinh \frac{JH_{l}V\cos \theta }{T}-2n^{+}\sinh \frac{JH_{l}V\cos
\theta }{T})]$
\end{center}

These equations can be rewritten in terms of the fractional magnetization $f=%
\frac{n^{+}-n^{-}}{n^{+}+n^{-}}$, becoming

\begin{center}
\bigskip $\frac{df}{dt}=2\omega e^{-\frac{KV}{T}}\cosh \frac{JVH_{l}\cos
\theta }{T}(\tanh \frac{JVH_{l}\cos \theta }{T}-f)$
\end{center}

\bigskip

At equilibrium \ $f=\tanh \frac{JVH_{l}\cos \theta }{T}$. At constant
temperature the differential equation is easily solved to yield

\begin{center}
$f=f_{0}e^{-(2\omega te^{-\frac{KV}{T}}\cosh \frac{JVH_{l}\cos \theta }{T}%
)}+\tanh \frac{JVH\cos \theta }{T}[1-e^{-(2\omega te^{-\frac{KV}{T}}\cosh 
\frac{JVH_{l}\cos \theta }{T})}]$
\end{center}

$\bigskip $

$f_{0}$ is the value of $f$ at $t=0$. The result shows that the initial
value of $f$ is decaying exponentially with time, and being replaced by the
equilibrium value at $T.$ Because of the double exponential, the time
dependence is strongly volume dependent, and if $JH_{l}<K$, or $H_{l}<H_{c}$
(the coercive force), to an excellent approximation if \ \ \ $V>\frac{T}{K}%
\ln \omega t,$\ $f=f_{0}$, and if$\ V<\frac{T}{K}\ln \omega t,$\ $f=\tanh 
\frac{JVH\cos \theta }{T}$, and

\ 

\begin{center}
$H_{l}=\lambda \int dVm_{l}(V)\cong \lambda \int_{0}^{\pi /2}\sin \theta
\cos \theta d\theta \lbrack \int_{0}^{\frac{T}{K}\ln \omega t}dVN(V)JV\tanh 
\frac{JVH_{l}\cos \theta }{T}+\int_{\frac{T}{K}\ln \omega t}^{V_{\max
}}dVN(V)JVf_{0}]$
\end{center}

\bigskip

During cooling, starting at a high enough temperature for all the grains to
be in equilibrium, they will initially be able to maintain their equilibrium
distribution. Eventually, however, the relaxation rate, $\frac{1}{\tau }%
=\omega e^{-\frac{KV}{T}}$, becomes too slow for this to be possible, and
the grains become blocked. The temperature at which this occurs, the
blocking temperature, is defined here as that temperature below which a
grain flips its magnetization no more than once, thus, if the rate of
cooling is $\frac{dT}{dt}=\alpha $, $\alpha \frac{dt}{\tau }$ flips will
occur in time $dt$, and the condition becomes $1=\int_{0}^{T_{b}}\frac{dT}{%
\alpha }\omega e^{-\frac{KV}{T}}=\frac{\omega }{\alpha }KV[\frac{T_{b}}{KV}%
E_{2}(\frac{KV}{T_{b}})]$ where $E_{2}(x)$ is the exponential integral
function of order $2$. Using the asymptotic expansion of $E_{2}$ [12], \ $%
\frac{\omega }{\alpha }\frac{T^{2}}{KV_{T}}e^{-\frac{KV}{T_{b}}}\cong 1$,
and $\frac{KV}{T_{b}}+\ln \frac{KV}{T_{b}}\cong \ln \frac{\omega }{\alpha }%
T_{b}$, \ $\omega \symbol{126}10^{8}$ Hz; so the value of $T_{b}$ does not
have much effect on the right hand side. Replacing $\frac{T_{b}}{\alpha }$
by a suitable average for the time it takes to cool the sample from $T_{b}$
to a low temperature, $\frac{KV_{b}}{T_{b}}+\ln \frac{KV_{b}}{T_{b}}\symbol{%
126}23$, and $\frac{KV_{b}}{T_{b}}\symbol{126}20$.

\bigskip

During cooling the magnetization at a temperature $T$ contains the sum of
the magnetizations of all grains whose blocking temperature is less than $T$%
, $JV\tanh \frac{JVH_{l}\cos \theta }{T}$, plus the magnetizations of all
the blocked grains, letting $x=\cos \theta $:

\begin{center}
\ \ \ \ \ \ \ \ \ \ \ \ \ \ \ \ \ \ \ \ \ \ \ \ \ \ \ \ $H_{l}\cong \lambda
\int_{0}^{1}xdx[\int_{0}^{V_{b}(T)}dVN(V)JV\tanh \frac{JVH_{l}(T)x}{T}%
+\int_{V_{b}(T)}^{V_{\max }}dVN(V)JV\tanh \frac{20JVH_{l}(T_{b})x}{T_{b}(V)}%
] $
\end{center}

\bigskip

If the sample has cooled to a temperature that is low enough for all the
grains to be blocked,

$\bigskip $

\begin{center}
$H_{l}\cong \lambda \int_{0}^{1}xdx\int_{0}^{V_{\max }}dVN(V)JV\tanh \frac{%
20JH_{l}(\frac{KV}{20})x}{K}$
\end{center}

If a small field, $h,$ is now applied at an angle $\phi $ to the grains, and
the sample heated to a temperature $T$, all grains whose blocking
temperature is less than $T$ will have the equilibrium distribution
corresponding to $T$\ .\ The magnetization will be

\ \ \ \ \ \ \ \ \ \ \ \ \ \ \ \ \ \ \ \ \ \ \ \ \ \ \ \ \ \ \ \ \ \ \ $%
M_{l}\cong \int_{0}^{1}xdx\{\int_{0}^{V_{b}(T)}dVN(V)JV\tanh \frac{%
JV[H_{l}(T)x+h\cos \phi ]}{T}+\int_{V_{b}(T)}^{V_{\max }}dVN(V)JV\tanh \frac{%
20JH_{l}(\frac{KV}{20})x}{K}]$

The local field is $\lambda M_{l}$; so heating results in a field equal to

\bigskip

\begin{center}
$H_{h}\cong $ $H_{l}+\int_{0}^{1}xdx\{\int_{0}^{V_{b}(T)}dVN(V)JV[\tanh 
\frac{JV(H_{l}x+h\cos \phi )}{T}-\tanh \frac{JVH_{l}x}{T_{b}(V)}]$.
\end{center}

\bigskip

If $T$ is low, $V_{b}(T)<<V_{\max }$, the integral is small and the
correction to $H_{l}$ can be neglected, and $H_{h}\approx H_{l}$.\ This will
be done in the following treatment.

\bigskip

Somewhere in the sample a region exists that is identical except that the
interaction field has the opposite sign:

\begin{center}
\bigskip

\ \ \ \ \ \ \ \ \ \ \ \ \ \ \ \ \ \ \ \ \ \ \ \ \ \ \ \ \ \ \ \ \ \ \ $%
M_{l}^{-}\cong \int_{0}^{1}xdx\{\int_{0}^{V_{b}(T)}dVN(V)JV\tanh \frac{%
JV[-H_{l}x+h\cos \phi ]}{T}-\int_{V_{b}(T)}^{V_{\max }}dVN(V)JV\tanh \frac{%
20JH_{l}x}{K}]$
\end{center}

\bigskip

Summing, and taking advantage of the smallness of $h$,

\bigskip

\begin{center}
$M\cong \frac{h\cos \phi }{T}\int_{0}^{1}xdx%
\int_{0}^{V_{b}(T)}dVN(V)(JV)^{2}\cosh ^{-2}\frac{JVH_{l}x}{T}$
\end{center}

\bigskip If the cooling is interrupted for a time $t_{w}$ at a temperature $%
T_{w}$, $H_{l}$ becomes

\begin{center}
$\ H_{l,w}\cong \lambda \int_{0}^{1}xdx[\int_{0}^{\frac{20T_{w}}{K}%
}dVN(V)JV\tanh \frac{20JH_{l,w}x}{K}+\int_{\frac{20T_{w}}{K}}^{\frac{T_{w}}{K%
}\ln \omega t_{w}}dVN(V)JV\tanh \frac{JVH_{l,w}x}{T_{w}}+\int_{\frac{T_{w}}{K%
}\ln \omega t_{w}}^{V_{\max }}dVN(V)JV\tanh \frac{20JH_{l,w}x}{K}]$

$=H_{l}+\Delta H=H_{l}+\lambda \int_{0}^{1}xdx[\int_{\frac{20T_{w}}{K}}^{%
\frac{T_{w}}{K}\ln \omega t_{w}}dVN(V)JV(\tanh \frac{JVH_{l,w}x}{T_{w}}%
-\tanh \frac{JVH_{l}x}{T_{b}(V)})]$

\bigskip
\end{center}

and the moment produced on heating is

\begin{center}
$M_{w}\cong \frac{h\cos \phi }{T}\int_{0}^{1}xdx%
\int_{0}^{V_{b}(T)}dVN(V)(JV)^{2}[1-\tanh ^{2}\frac{JV(H_{l}+\Delta H)x}{T}]$
\end{center}

$\bigskip $

$\Delta H$ is clearly small; so

$\bigskip $

\begin{center}
$\Delta M=M_{w}-M\cong -2\frac{J^{3}}{T^{3}}h\cos \phi
\int_{0}^{1}x^{2}dx\int_{0}^{V_{b}(T)}V^{3}dVN(V)[\tanh \frac{JVH_{l}x}{T}%
\cosh ^{-2}\frac{JVH_{l}x}{T}]\Delta H$
\end{center}

The quantity $\tanh \frac{JVH_{l}x}{T}\cosh ^{-2}\frac{JVH_{l}x}{T}$ reaches
a maximum at $\frac{JVH_{l}x}{T}\approx 0.6$, and drops to about 10\% of its
maximum when $\frac{JVH_{l}x}{T}\approx 1.5$. Therefore it can be concluded
that if the local field becomes very large the effect of waiting will
disappear, and the spin glass is replaced by a magnetic glass. In this
connection no ageing effects at all were observed in two different
concentrated assemblies of cobalt particles, one with a mean size of 3 nm,
the other with a mean size of 8 nm [13]. The samples were prepared by
evaporating the liquid from a colloidal suspension to dryness; so the
individual particles were separated by the thickness of a 2nm polymer layer.
In both cases the shift in the position of the peak in the ZFC magnetization
was substantial, with the temperature of the peak more than doubling,
suggesting that the local field was on the order of the coercive force.

\bigskip

Bearing the above considerations in mind the remaining analysis assumes $%
\frac{JVH_{l}x}{T}<1$, in which case,

$\bigskip $

\begin{center}
$\Delta M\approx -\frac{J^{4}H_{l}}{6T^{4}}h\Delta
H\int_{0}^{V_{b}(T)}V^{4}dVN(V)$
\end{center}

$N(V)$ is usually approximated by a log-normal distribution: $%
N(V)\thickapprox \frac{e^{-\frac{\ln V-\ln V_{m}}{2\sigma ^{2}}}}{V}$ where $%
\sigma $ is the standard deviation, and $V_{m}$ is the most probable volume;
so, letting $b=\sqrt{2}\sigma $, and $\frac{V}{V_{m}}=x$ \ 

$\bigskip $

\begin{center}
$\Delta M=-\frac{hH_{l}\Delta H}{6}(\frac{JV_{m}}{T})^{4}\int_{0}^{\frac{T}{%
KV_{m}}\ln \frac{\omega T}{20\beta }}x^{3}dxe^{-\frac{1}{b^{2}}\ln ^{2}x}$

$\bigskip $Now $KV_{m}\symbol{126}20T_{m}$ where $T_{m}$\ is the temperature
of the peak in the ZFC\ magnetization curve for a dilute \ \textit{%
non-interacting }sample; so

$\Delta M=-\frac{hH_{l}\Delta H}{6}(\frac{JV_{m}}{T})^{4}\int_{-\infty
}^{\ln \frac{T}{T_{m}}}e^{3y}e^{-\frac{y^{2}}{b^{2}}}=-\frac{\sqrt{\pi }%
hH_{l}\Delta H}{12}(\frac{JV_{m}}{T})^{4}e^{\frac{3b^{2}}{4}}[1+\func{erf}(%
\frac{\ln \frac{T}{T_{m}}-\frac{3b^{2}}{2}}{b})]$
\end{center}

letting $\gamma =\frac{T_{w}}{T_{m}},y=\frac{T}{T_{w}}$ , $\Delta M$ $\ $can
be rewritten as $\ \Delta M\ \thickapprox \frac{const.}{y^{4}}\func{erf}c(%
\frac{3}{2}b-\frac{\ln \gamma y}{b})$

The resulting temperature dependence is very strong, and the magnitude of $%
\Delta M$ \ increases rapidly with temperature. This is illustrated in the
figure below. However when the temperature exceeds $T_{w}$ the effect of
waiting begins to be removed, since the grains affected by the delay are now
having their moments reset. If $\Delta H_{0}$ is the value for $T<T_{w}$,
when $T>T_{w}$

\begin{center}
$\Delta H\simeq \Delta H_{0}\frac{\int_{\frac{T}{T_{m}}}^{\frac{aT_{w}}{T_{m}%
}}xdxe^{-\frac{1}{\sigma ^{2}}\ln ^{2}x}(\frac{1}{T}-\frac{1}{T_{m}x})}{%
\int_{\frac{T_{w}}{T_{m}}}^{\frac{aT_{w}}{T_{m}}}xdxe^{-\frac{1}{\sigma ^{2}}%
\ln ^{2}x}(\frac{1}{T_{w}}-\frac{1}{T_{m}x})}=\Delta H_{0}\frac{\frac{1}{y}[%
\func{erf}(\frac{\ln a\gamma -\sigma ^{2}}{\sigma })-\func{erf}(\frac{\ln
\gamma y-\sigma ^{2}}{\sigma })]-\gamma e^{-\frac{3\sigma ^{2}}{4}}[\func{erf%
}(\frac{\ln a\gamma -\frac{\sigma ^{2}}{4}}{\sigma })-\func{erf}(\frac{\ln
\gamma y-\frac{\sigma ^{2}}{4}}{\sigma })]}{[\func{erf}(\frac{\ln a\gamma
-\sigma ^{2}}{\sigma })-\func{erf}(\frac{\ln \gamma -\sigma ^{2}}{\sigma }%
)]-\gamma e^{-\frac{3\sigma ^{2}}{4}}[\func{erf}(\frac{\ln a\gamma -\frac{%
\sigma ^{2}}{4}}{\sigma })-\func{erf}(\frac{\ln \gamma -\frac{\sigma ^{2}}{4}%
}{\sigma })]}$

\bigskip
\end{center}

and $\Delta H$, hence $\Delta M$ rapidly disappear at temperatures above $%
T_{w}$. $\Delta M$ is plotted in the figure.

\bigskip

Considering the limitations of a mean field approach the agreement between
theory and experiment is excellent. The failure to reproduce the rounded
transition from decreasing to increasing $\Delta M$ at $T_{w}$ can be traced
to the initial assumption that grains block abruptly at the blocking
temperature.

\bigskip

The strong dependence of the effect on the width of the grain size
distribution is interesting, but not surprising: rejuvenation is due to the
interaction field frozen in by the grains blocked on cooling. Waiting allows
grains that were blocked during cooling to $T_{w}$ to reach their
equilibrium distribution at $T_{w}$, and the wider the grain size
distribution the wider will be the temperature range over which the grains
grains block. "Rejuvenation" is just the gradual replacement of the
interaction field frozen in during waiting by the equilibrium field during
heating; so, the wider the distribution, the larger the number of smaller
grains that will reach their equilibrium distribution during waiting, and
become blocked on cooling.

\bigskip

It has been shown that a straight forward mean field theory can account for
rejuvenation in superspin glasses. In contrast to other approaches no
droplets are assumed to exist, nor is a hierarchy of energy levels invoked,
and while a chaotic distribution of spins is consistent with the theory, no
aspects of chaos theory are employed.

\bigskip

\bigskip

\begin{center}
\textbf{References}
\end{center}

\bigskip 1) D. Walton, Nature 286, 245 (1980); Nature 305, 616 (1983)

2) C. Djurberg et al., Phys. Rev. Lett. 79, 5154 (1997\bigskip )

\bigskip 3) T.Jonsson et al. Phys. Rev. Lett. 75, 4138 (1995)

4) H.Mamiya, I.Nakatani and T.Furubayashi, Phys. Rev. Lett. 82, 4332 (1999)

5) K.Jonasson et al., Phys. Rev. Lett. 81, 3243 (1998)

6)  T.Jonsson et al., Phys. Rev.B 59, 8770 (1999)

7)  C.Djurberg,K.Jonason and P.Nordblad, Eur. Phys. J.B 10, 15 (1999)

\bigskip 8) H.Mamiya and I.Nakatani, Nanostr. Mater. 12, 859 (1999)

9) T.Jonsson, M.F.Hansen, and P.Nordblad, Phys. Rev. B 61, 1261 (2000)

10) S.Sasoo et al.,Phys.Rev.B 65, 134406 (2002)

11) M.Sasaki et al.,Phys.Rev.B 71, 104405 (2005)

12) M. Abramowitz and I. A. Stegum, "Handbook of Mathematical Functions",
Dover, New York, N.Y.

13) D. Walton and N.Thanh, unpublished.

\bigskip

\bigskip

\bigskip

\FRAME{ftbpF}{5.3287in}{4.0053in}{0in}{}{}{Figure}{\special{language
"Scientific Word";type "GRAPHIC";maintain-aspect-ratio TRUE;display
"USEDEF";valid_file "T";width 5.3287in;height 4.0053in;depth
0in;original-width 7.6255in;original-height 5.7224in;cropleft "0";croptop
"1";cropright "1";cropbottom "0";tempfilename
'J2M2T900.wmf';tempfile-properties "XPR";}}

\bigskip

Figure 1: the difference in moment introduced on heating between two samples
cooled in zero field in which one was held at an intermediate temperature
normalized to the difference in the moment at the waiting temperature
plotted against the temperature normalized to the waiting temperature. The
three curves on the left of the figure are computed for 3 different values
of the standard deviation in a log-normal particle size distribution. The
value of the standard deviation had negligible effect on the curve on the
right of the figure. The open circles are experimental data taken from
Sasaki et al. [11].

\end{document}